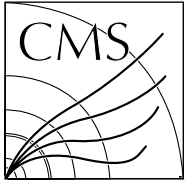

**The Compact Muon Solenoid Experiment**

# CMS Note

Mailing address: CMS CERN, CH-1211 GENEVA 23, Switzerland

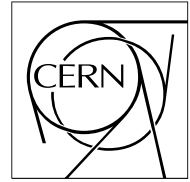

# Object Serialization and Deserialization Using XML

J.-M. Le Goff, H. Stockinger, I. Willers

*CERN, Geneva, Switzerland*

R. McClatchey, Z. Kovacs, P. Martin

*Centre for Complex Cooperative Systems, Univ. West of England, Frenchay, Bristol BS16 1QY UK*

N. Bhatti, W. Hassan

*Computer Training Centre, Islamabad, Pakistan*

### Abstract

Interoperability of potentially heterogeneous databases has been an ongoing research issue for a number of years in the database community. With the trend towards globalization of data location and data access and the consequent requirement for the coexistence of new data stores with legacy systems, the cooperation and data interchange between data repositories has become increasingly important. The emergence of the eXtensible Markup Language (XML) as a database independent representation for data offers a suitable mechanism for transporting data between repositories. This paper describes a research activity within a group at CERN (called CMS) towards identifying and implementing database serialization and deserialization methods that can be used to replicate or migrate objects across the network between CERN and worldwide centres using XML to serialize the contents of multiple objects resident in object-oriented databases.

# 1. Introduction

The Compact Muon Solenoid, CMS [1], a physics experiment at CERN, the European Organization for Nuclear Research, has a large number of Terabyte sized databases already being generated during its construction phase. The final assembly and installation of the CMS detector will take place at CERN and the experiment is due to take its first data in mid 2005 after which a Petabyte of raw event data will be generated and stored per year. The CMS construction and simulation databases now being produced use the Objectivity/DB [2] object database system. At CERN there is a need to exchange data among repositories both at the construction level (for transfer of parts information between the local centers and also to the central system at CERN) and for physics analysis data (for transfer of physics data for analysis to the regional centers). At present the only way to transfer this data between repositories is manual and database- and schema-dependent. The contribution of this paper is to provide the database community such a tool using Objectivity/DB and to demonstrate the feasibility of the approach through real-world applications. The diversity of sites and software used makes it essential to provide tools for efficient data conversion and data transfer in a distributed computing environment. XML is an agreed standard for information exchange and allows the gap between distributed data sources to be bridged.

The CMS detector consists of seven major systems called subdetectors: the magnet, tracker, electromagnetic calorimeter, hadron calorimeter, muon detector, trigger and data acquisition. There will also be infrastructure including services such as electricity, cooling and ventilation. As these complex systems are manufactured and assembled, databases need to be built up that describe individual items and their assembly. Data must be made available across potentially heterogeneous databases to satisfy the data management requirements of the CMS subdetectors. As one example, the Electromagnetic CALorimeter, or ECAL, group has developed specialist software called CRISTAL (see later) to keep track of its construction and assembly information. The CMS subdetectors are manufactured and assembled in various centres around the world. The CRISTAL database for ECAL is distributed between centres in Russia, China, France and Switzerland. The main construction database is at CERN with objects being sent to other centres where additional information is collected and transferred back to CERN in the form of objects. Recently it was decided to explore the use of the new eXtensible Markup Language, XML, standard for a number of applications including object transfer between centres.

A number of computing centres in CERN, Pakistan and UK (see author list) have formed a project called WISDOM [3] (Wide-area database-Independent Serialization of Distributed Objects for Migration) in order to construct software that investigates the use of the XML [4] standard. The aim of the first phase is to acquire the ability to serialize Objectivity/DB objects, turn the serialized objects into XML, and deserialize the objects, turning XML into Objectivity/DB database objects. The WISDOM project deals with the exchange of data among multiple database inter-linked through a wide-area network in a database independent way by using XML. XML brings the following advantages:

- It supports transfer over the network.
- It is a human 'readable' and an easily 'understandable' format.
- It is language- and platform-independent.
- The validity of the XML document, after transfer over network, can easily be checked at the destination using a standard parser.

The process of collecting the complex data structures from the database in XML format is called serialization. Similarly the process of recreating the complex structure back from the XML format is called deserialization. The main purpose of XML serialization is to move selected objects (including collections) between persistent stores. Serialization is employed to provide coherent duplication of consistent datasets.

In WISDOM a C++ class has been designed for serialization which is responsible for serializing the complex data "objects" and its model, the "schema". Two separate XML files are generated by the Serializer class (called the SchemaXML and ObjectXML files) and these files can be exchanged between the repositories. Once the files reach the destination they are handled by the deserializer class for the re-creation of the objects. The deserializer first recreates the schema and then populates the schema with objects. The WISDOM project is making the process of replication automatic and database independent. In the final phase this project will provide a tool for managing the transfer of objects between repositories in a database-independent way. This tool is designed in Java2 and is consequently platform-independent.

The layout of this paper is as follows. In the next section the background to serialization and deserialization in the WISDOM project is introduced and this is compared to related work in the following section. The implemented serialization and deserialization tools are then described. Applications of object (de-)serialization and future work are later discussed and their use in CMS is outlined.

## 2. Object Serialization and Objectivity/DB

The decomposition of complex data structures such as Objectivity/DB data containers, into a sequence of their primitive data parts, which can be saved directly in a file or transferred over a network, is referred to as serialization. Object serialization means storing an objects state and model in a form that could be accessed serially, such as storing the object in a disk file, or transferring the object via a network, such as with HTML data. The serialized data are used to reproduce or re-create the object. Every object has a state, including the current values stored in various member variables of the object and its model, i.e. a skeleton of the object with member variable and member function definitions. To serialize an object, both the state and model of the object must be serialized.

The re-creation of complex data structures or data containers, from a sequence of their primitive data parts is called de-serialization. Deserialization implies re-creating the object by reading its state/model from some serialized data. The underlying object model plays an important role in the serialization/de-serialization process, as it gives complete information about the object. The object model defines the names of the member variables associated to the object, the possible operations defined in the object and any links of the object with other objects. This study reports on the serialization and de-serialization of objects from an Objectivity/DB database. The objects created in an Objectivity/DB base are persistent. Objects have two parts: 'The Object Data' and 'The Object/Data Model', called the Schema. With Objectivity/DB the serialization of object data is relatively straightforward as Objectivity/DB provides methods to open an object from a database and access its data and other attributes such as its type and its unique Object ID (OID).

The general information about the object (e.g. OID and type) is accessible to every application while the access to object data is possible only if its schema is known. Knowing the schema, a Serializer application can place this information in an appropriate format for a Deserializer. This serialized object can be transported to a Deserializer application as a serialized object. The serialized object can then be de-serialized by re-creating it at the destination database and accessing the values stored in the serialized object. Clearly, this re-creation can only be possible if the schema of the object is known to the De-Serializer application as well as the Serializer application. This can only be achieved by serializing the object model (the schema) along with the object data, and by this schema being made known to the database at the destination.

A schema-independent application can be developed that opens any object, in any Objectivity/DB container in any database of any federation and can write its serialized object. (Hereafter these expressions are used as Objectivity-specific definitions). In this work a Serializer/ Deserializer application has been developed that can be used to replicate/migrate an Objectivity/DB database for an application-specific domain.

## 3. Related Work

Any serialized representation of an object should have the following capabilities:

- It should be platform and language independent, since serialization and deserialization could be carried out on different platforms.
- Its validity must be easily verified.
- It should be simple to deserialize.

Currently there is much effort going on in using XML as a means of serializing objects. The following research areas can be distinguished: serializing Java objects, serializing data from relational databases into XML and serializing persistent objects from object-oriented databases. One example is KOML, the KOala xML serialization tool [5] that provides an easy way to serialize and deserialize any Java objects in an XML document. KOML is similar to the approach adopted in this paper in that two main classes for serialization and deserialization are provided. KOML takes advantage of the built-in streaming features of the Java language, which is an important difference between KOML and WISDOM, since streaming cannot be used in WISDOM's C++-based approach. However, KOML does not have any binding for an object-oriented database management system such as Objectivity/DB.

The Casbah distributed objects system allows for different object serialization formats that are self-describing [6]. The project uses features of the XML specification for serialization purposes. XML Serialization defines three generic datatypes, <atom>, <list>, and <dictionary> that can be further specialized by higher level protocols or marshalling. Since there is no database binding, the project is not directly related to our work. Sun is also currently developing a tool called the XML Data-Binding facility which provides an automatic translation between XML and Java objects [7]. This is related to the work presented in this paper, except that here persistent C++ objects are serialized.

The XML Metadata Object Persistence (XMOP) allows interoperation between object technologies such as Java,

Microsoft COM and CORBA [8]. XMOP is unique in currently available object serialization in that it is not directly tied to a particular object system. Currently there is no effort going on to serialize objects into XML created with Objectivity/DB. This paper redresses this deficiency.

## 4. Software Reuse

In the serialization philosophy adopted for the WISDOM project, to enable migration of data between databases it is necessary to convert the database schema into a suitable format for transfer. Without the capability of the so-called Active Schema facility in Objectivity/DB any implementation of the serializer would, by definition, become schema dependent. Hard-coding the schema information, however, makes the Serializer schema-dependent. With the newly released Active Schema API an application can retrieve schema information dynamically and consequently schema-independence can be achieved.

The latest release of Objectivity/DB, Objectivity 5.2, includes this Active Schema (AS) facility [2]. The AS facility allows programmers to dynamically interrogate the database schema. Among many capabilities, the AS provides facilities for obtaining class descriptions and modifying class specifications dynamically. Programmers have access to domain descriptions at runtime, consequently allowing them to access and manipulate the schema. The AS facility is a tool which can help in the creation of a generic query facility for Objectivity/DB databases (see later example). A generic query facility for Objectivity/DB databases can use the AS facility to load and interrogate the database schema dynamically, thereby not restricting the query facility to a particular domain specification.

The WISDOM project is using XML as the data format for serialized objects for the following reasons:

- The XML document (DTD) contains the description of the primitive data parts of the objects i.e. it contains the 'data' as well as their 'context'.

- Standard parsers/de-parsers are available for the deserializing application to parse and deparse the DTD.

Furthermore, XML is an appropriate format for moving data between databases (object-oriented and/or relational) since it supports the separation between the semantic and graphical representation of the data and the data instances themselves. This facilitates the separation of the description of the database architecture from the description of the database schema, leading naturally to a generic architecture for moving data between heterogeneous databases. Using these basic ideas and the Active Schema API, a schema independent Serializer/Deserializer has been developed that can be used to serialize any Objectivity/DB Database. The Serializer is provided as a C++ class that can be used by any C++ Objectivity/DB application to serialize any object from the database.

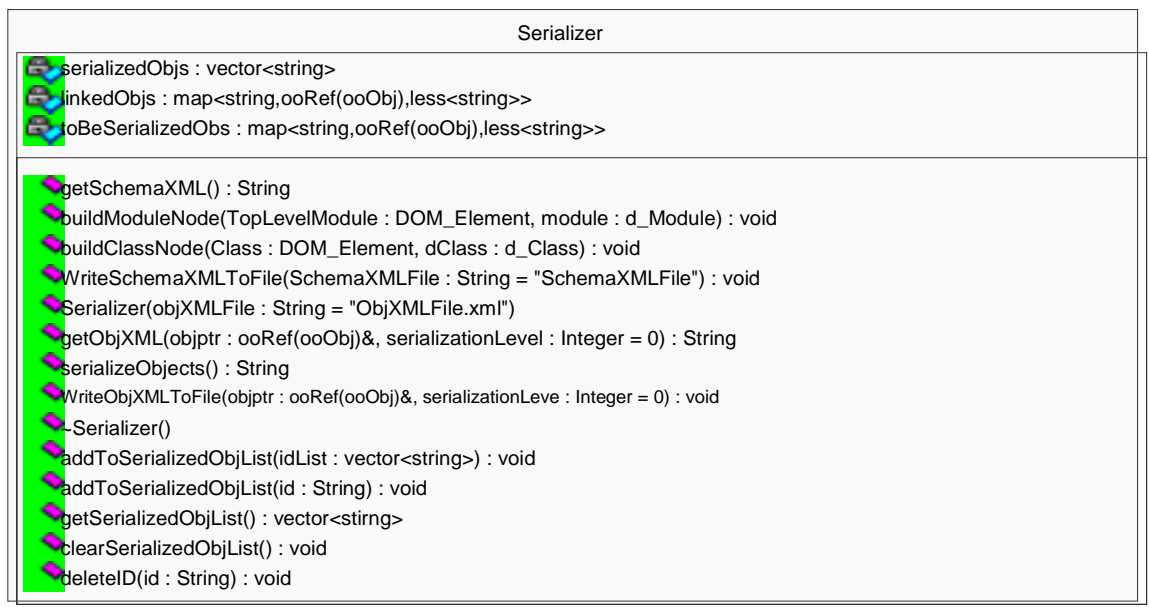

**Figure 1. UML model of the Serializer.**

# 5. The Implemented Serializer

## 5.1 The Serializer Class

Figure 1 shows the UML [9] model of the Serializer class. The Serializer class acts as the interface to the User/Application (class(es)) and also as a "coordinator" among its various "component" classes. The Serializer uses the AS API to dynamically read the schema and object information from an Objectivity/DB federation. This dynamically read information is packed into the XML Document Object Model (DOM) [10] tree using the DOM API. This DOM tree is then dumped into a stream using an overloaded operator for DOM elements. The stream can be connected to a buffer in memory or to a file on the disk.

Logically the interface of the Serializer has three sections through which the applications can use the Serializer: Schema XML Generation, Object XML Generation and Serialized Object List Maintenance. Once both the serialized schema and object XML files are created and transferred to the destination, there is a need to recreate the schema and all the states of the serialized objects. In the process of serialization the persistent objects are transferred from the Objectivity/DB to another persistent layer i.e. an XML layer. At this point these files must be interpreted to get the object states and models back to the base. That is deserialization takes place.

**DDL:**
```
class APersistanceClass : public ooObj
{
 public:
//Constructor
       APersistanceClass(int);
       APersistanceClass(){}
//Associations
//Bidirectional
//One to One
    ooRef(DerivedClass) derivedClass_OneToOne <-> aPersistanceClass_OneToOne:copy(delete),version(delete);
    ooRef(BaseClass1) baseClass1_OneToOne <-> aPersistanceClass_OneToOne:copy(delete),version(delete);

//Destructor
       ~APersistanceClass(){}
//Interface
private:
       int16 DC_BasicAttribute;
        ooVString DC_ooVString;
};
```

**SchemaXML:**
```xml
<Schema>
<TopLevelModule>
<Class Name="APersistanceClass" TypeNumber="1000000">
   <BaseClass Name="ooObj" Position="0" AccessKind="PUBLIC" />

         <Relationships>
<Bidirectional Name="derivedClass_OneToOne" IsShort="false" CopyMode="CopyDrop" IsInline="false" IsToMany="false" Position="1" AccessKind="PUBLIC" Versioning="VersionDrop" InverseName="aPersistanceClass_OneToOne" Propagation="NoInfo" OtherClassName="DerivedClass" InverseIsToMany="false" SpecifiedAssocNum="3020898880" />

<Bidirectional Name="baseClass1_OneToOne" IsShort="false" CopyMode="CopyDrop" IsInline="false" IsToMany="false" Position="2" AccessKind="PUBLIC" Versioning="VersionDrop" InverseName="aPersistanceClass_OneToOne" Propagation="NoInfo" OtherClassName="BaseClass1" InverseIsToMany="false" SpecifiedAssocNum="873415233" />
</Relationships>
<Attributes>
<BasicAttribute Name="DC_BasicAttribute" Type="int16" Position="9" ArraySize="1" AccessKind="PRIVATE" HasDefaultValue="false" />
<EmbeddedClassAttribute Name="DC_ooVString" Position="10" ArraySize="1" AccessKind="PRIVATE" OtherClassName="ooVString" />
</Attributes>
</Class>
</TopLevelModule>
    </Schema>
```

**Figure 2. Objectivity DDL and the associated schemaXML file.**

The Schema XML Generation section consists of the following functions (see figure 1):

*string getSchemaXML()*
*void writeSchemaToXMLFile(char * schemaXMLFile= "SchemaXMLFile.xml")*

The first function returns a string, containing the schema in an XML format. The second function uses the first function to get the XML string and writes it to a file called schemaXML.

Figure 2 shows an example of some DDL from Objectivity/DB and the corresponding schemaXML file produced by the Serializer. The Object XML Generation section consists of the following functions:

*string getObjXML(ooRef(ooObj)& objptr, int serializationLevel)*
*void writeObjToXMLFile(ooRef(ooObj)& objptr, int serializationLevel)*

Here the first function serializes the specified object and the objects linked to it, up to a specified serializationLevel, and returns the generated XML string. If the serializationLevel is set to a negative value, the whole web of objects connected to the specified object is serialized.

The Serialized Object List Maintenance section consists of the following functions:

*void addToSerializedObjList(vector<string>& idList)*
*void addToSerializedObjList(string id)*
*void clearSerializedObjList()*
*void deleteID(string id)*

After serializing an object, the serializer adds its ID to the SerializedObjList vector. Before serializing an object, it checks the SerializedObjList to see if the object is already serialized or not. If not, it serializes the specified object. An application using the Serializer can directly interact with the SerializedObjList vector using the Serialized Objects List Maintenance section of the interface, which results in some interesting usage scenarios (see later).

## 5.2 A Modified Overloaded Operator for DOM Nodes

The standard operator overloading function has the ability to dump a DOM node and its child nodes in an XML format regardless of the fact that the node is a Document node, a Document Type node or an ordinary Element node. This has the drawback that the whole DOM structure has to be created in memory prior to dumping it into a file which is impractical for large DOM trees due to memory shortage.

In WISDOM this standard behaviour has been modified so that the operator prints <?xml version='1.0' encoding='utf-8' ?> in case of Document nodes. For Document Type nodes the function checks if the XML is being generated for a Schema or for Objects and includes the respective DTD in the DOCTYPE element. The behaviour is the same for ordinary Element nodes i.e. dump all the nodes in the stream which assists in generating XML for large databases.

For objects the XML generation has the following sequence;

1. Firstly, in the current Serializer the objects constructor prints <?xml version='1.0' encoding='utf-8' ?>, DOCTYPE node together with the starting tag of the root node in the XML file and this file is referred to as the objectXML file.

2. Any request to serialize an object results in an Object element node. This node is dumped to the objectXML file.

3. At the end of the scope of the Serializer object, the destructor prints the closing tag of the root element in the XML file and closes the objectXML file.

## 5.3 Usage scenarios

The currently implemented XML Serializer can be used to generate XML either in a memory buffer or in a file on the disk in the following ways:

- Generate Schema XML
- Generate the XML of an object.

- Generate the XML of an object and the objects linked to it, up to a specified level
- Generate the XML of an object and the complete web of the objects linked to it.
- Generate the XML of all objects linked to a specified object.
- Generate the XML of a subset of objects linked to a specified object.

Different combinations of the following input parameters make all the above permutations possible:

- Serialization level
- Object to be serialized
- Objects not to be serialized

# 6. The Implemented Deserializer

## 6.1 The Deserializer Class

The Deserializer class has been developed with generalization in mind: any applications that need to deserialize an object from any XML input can use the Deserializer. The Deserializer uses the Active Schema API to dynamically read and create the schema as well as objects information in an Objectivity/DB federation. This dynamically read information is used by the Deserializer to populate the database, getting object info from the serialized objects file. Its UML model is shown in Figure 3.

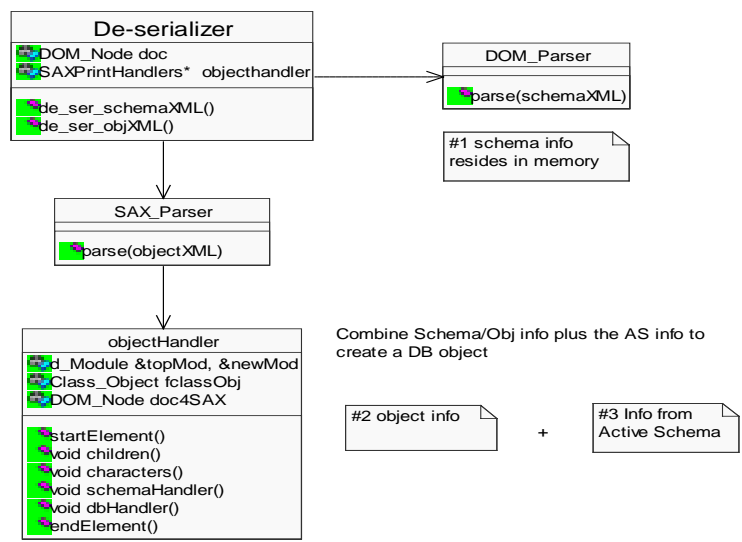

**Figure 3. UML model of the Deserializer**

The Deserializer uses the Objectivity Active Schema to recreate the schema at the destination and this schema is later used to populate the DB at the destination.

The interface comprises the following two methods for Schema and Object Deserialization:

*De_serialize_schema(schemaXMLFile.xml)*
*De_serialize_obj(objectXMLFile.xml)*

The Schema Deserialization section gets the schemaXML file (created by the Serializer see figure 2) from the application using the Deserializer. The file is sent to the DOM parser object, which creates the memory-resident tree structure of the schema. This allows the Deserializer to have rapid access to the schema thoughout its lifetime.

The Object Deserialization section gets the objectXML file and passes the file to a SAXParser (for an example of an objectXML file, corresponding to the DDL of figure 2, see figure 4). The use of the SAXParser ensures that if the size of the objectXML file is too large then the program does not crash. The SAXParser parses the objectXML file tag by tag and makes decisions either to create the object at parse time or to store information for deferred creation. The following function in the SAXparser is of principal interest:

*void startElement(const XMLCh\* const name, AttributeList& attributes);*

(Here "name" is the name of the XML tag). This function is called as soon as a tag is reached in the objectXML file. The SAXparser traverses the objectXML file serially. At the outset of parsing information is gathered about each object i.e its Objectivity database, container, name and typenumber etc. and after that the Deserializer attempts to resolve the class in the database schema. If the class is not resolved then the built-in DOM structure of the schema is searched for the particular object and the schema of the class is dynamically proposed and created by the AS API by calling the following function:

*void children(DOM_Node parent, Proposed_Class &factory);*

Note that the proposed class reference is also passed to the function. If the schema is present but there are some inconsistencies between the source and the destination schema, then the schema is evolved according to the source schema. This functionality is also possible because of the use of the Active Schema API.

**ObjectXML:**
```
<?xml version="1.0" encoding="utf-8" ?>
<!DOCTYPE ObjectList (View Source for full doctype...)>
<ObjectList>
<Object id="2-2-3-1" typename="APersistanceClass" typnumber="1000000">
<Database id="2-0-0-0" name="ooDefaultDB" typename="ooDBObj" typnumber="1004" />
<Container id="2-2-1-1" name="_ooDefaultContObj" typename="ooDefaultContObj" typnumber="1005" />

        <Relations>
<ToOne name="derivedClass_OneToOne">
<ID>2-2-3-7</ID>
</ToOne>
<ToOne name="baseClass1_OneToOne">
<ID>2-2-3-4</ID>
</ToOne>
        </Relations>

        <Attributes>
<Basic name="DC_BasicAttribute" type="int16">
<BasicElement index="0">10</BasicElement>
</Basic>
<String name="DC_ooVString" type="ooVString">
<StringElement index="0">ooVString</StringElement>
</String>
        </Attributes>

</Object>
    </ObjectList>
```

**Figure 4. An objectXML file for the DDL example shown in figure 2.**

Once the schema has been populated, control is transferred back to the SAXParser and it parses the other incoming tags. (Logically the next tags are either the "Attribute tag" or "Relationship tag"). At this point the actual data present in the objectXML file are collected by the following function:

*void characters(const XMLCh\* const chars , const unsigned int length)*

After collecting the actual data this function, in turn, calls the following function for populating the database:

*void dbHandler();*

Here "dbHandler" is a function responsible for the actual population of the data in the database. It gets the information from the following:

- The objectXML file
- The schema DOM structure in the memory or
- Dynamically from the schema through AS API.

## 6.2 Usage Scenarios

The currently implemented XML Deserializer can be used to populate the database in the following ways:

- Populate the Schema XML
- Populate the XML of an object.
- Populate the XML of an object and the objects linked to it, up to a specified level
- Populate the XML of an object and the whole web of the objects linked to it.

## 7. XML-Based Data Interchange Between CMS Databases

This section describes examples where the Serialization/Deserialization tools of WISDOM are relevant to the CMS experiment at CERN, and advantages gained from the XML de/serialization tool are identified . In general, these applications are characterized by a highly distributed and heterogeneous environment and the need to move data between databases in a database-independent format.

## 7.1 The CRISTAL Project

The construction of the CMS detectors is a complex process taking place in multiple laboratories or institutes located worldwide, over long timescales and it significantly extends understanding in engineering processes, in computer science and in physics. The CMS physicists require that the position, production process and characteristics of each detector part be captured in a database system. The CRISTAL project [11], [12] manages the production over geographically separated manufacturing centres of CMS components. A detailed description of this can be found in [13]. In summary CRISTAL has a single Central System which manages a collection of distributed manufacturing centres, each running versions of defined production processes and each gathering up to a Terabyte of construction-specific data.

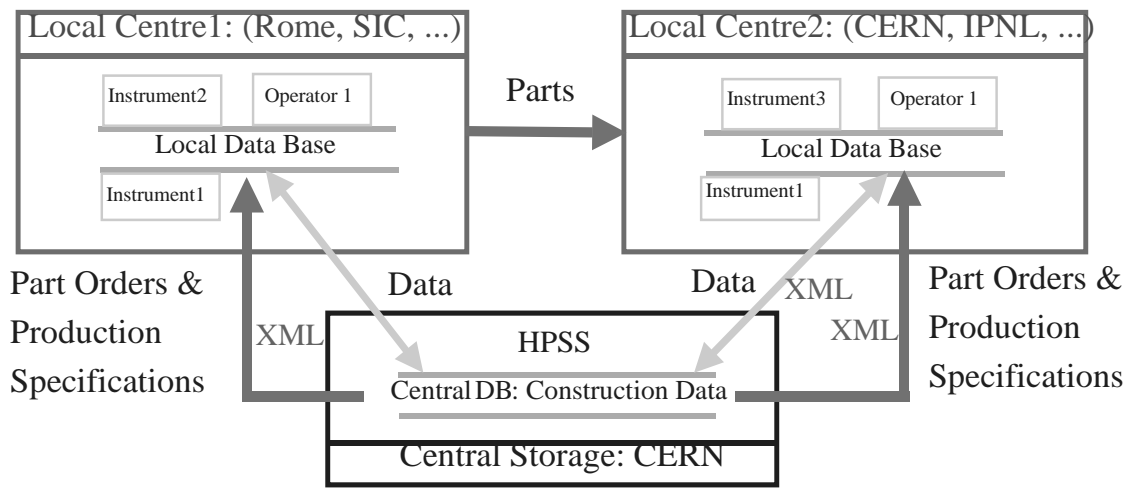

**Figure 5. The movement of data between CRISTAL centres using XML.**

In the CRISTAL Central System physicists specify what is to be built and how it is to be built, using a model which spans design to production. The physicists create product types, activity types and outcome data types and this information is distributed to all the manufacturing centres. The design of the CRISTAL prototype was dictated by the requirements for adaptability over extended timescales, for system evolution, for interoperability and for complexity handling and reusability. In adopting a description-driven design approach to address these requirements, a separation of object instances from object descriptions instances was needed. This abstraction resulted in the delivery of a two layer architecture - a model plus an associated meta-model, described in [13].

The system architecture within a manufacturing centre, like most modern distributed systems, is 3-tier as opposed to monolithic or simple client/server. Java user interfaces (thin clients) provide workflow interfaces to operators and access C++ CORBA server objects, which contain the business logic of the system. Business logic in CRISTAL is the interpretation, execution and management of workflows. These server objects in turn access an Objectivity/DB database through a C++ binding.

In each distributed manufacturing centre, data are collected from instruments and transferred from the instrument to the local Objectivity/DB in XML format. Each centre is autonomous and continues to collect data even if the link to the Central System is unavailable. When availability allows, the database is transferred, serialized, in an XML format between the remote manufacturing centres and the CERN-based Central System (see Figure 5). Using serialized XML provides for standardised data transfer and allows the use of industry-provided parsers/de-parsers for XML generation and interpretation. In addition, use of XML provides database-independence so that data can be moved to and from different database implementations.

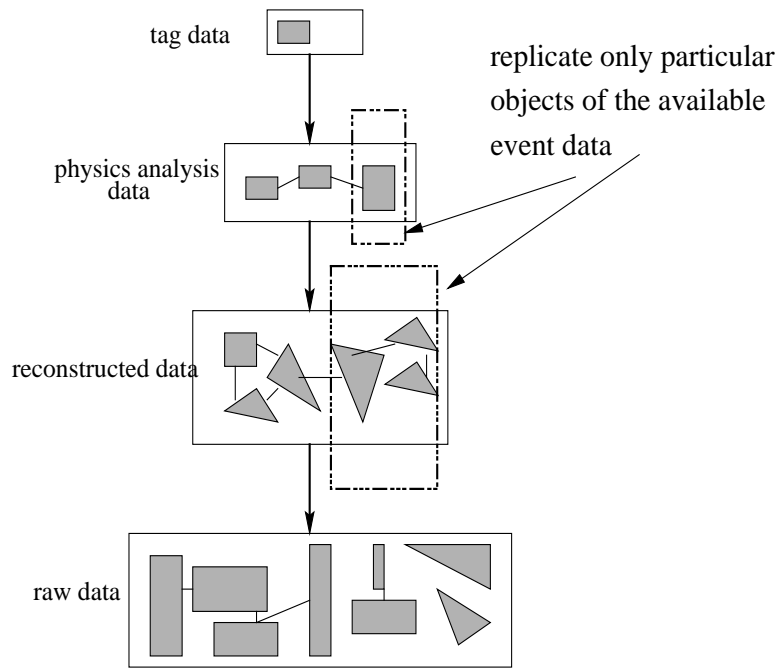

**Figure 6. Replication of objects using XML in the CMS Event Data Model**

## 7.2 The CMS Event Data Model

The current CMS event data model foresees different types of data being collected by the completed CMS detector from 2005 onwards. Initially data which are produced by the detector have to go through a trigger system where data are filtered by dedicated hardware and software. This is done to reduce the amount of data to be stored. The data written after the trigger system are called raw data or events and amount to over a Petabyte of data per year.

Basically, there is a hierarchy of objects to be stored in a large event object store. The raw data event is the lowest level in the hierarchy. A reconstruction function takes raw events as input, produces new objects called reconstructed event data objects and puts the newly created objects back into the object store. Further analyses of reconstructed objects can be carried on demand. The objects produced can also be called event summary data or analysis object data. These are the actual objects which are used for final analysis by the physicists. The smallest data type that can be distinguished is the so-called 'tag data' that stores summary information about raw, reconstructed, event summary and raw data objects.

Currently Objectivity/DB is deployed for storing event data objects [14]. The smallest unit of physically stored objects is an Objectivity/DB database that is mapped to a physical file. Furthermore, the smallest granularity for replication is such a file [15]. Assuming that the file size will be about 2 GBs, there will be significant network traffic for transferring single files over the network. For certain kinds of physics analysis it is assumed that only small parts of the file are interesting to the physicist.

It is not intended to use XML as a means of bulk data transfer between two physics analysis sites (see Figure 6). It is clear that ASCII (or even unicode) data has much more overhead than binary data stored in an object database. For example, consider objects of size 800 bytes (a class with eight VArrays of 100 uint8). If only 10 of these objects (8.000 bytes) have to be transferred to a remote site it is more efficient in terms of network traffic to do this with an XML file. Note that the XML file will have a certain overhead for metadata that describes the data objects. It still has to be studied when the overhead of an Objectivity/DB database file is bigger than that of an XML file. Only when a small number of objects need to be replicated, is there a case for dumping object data into an XML file. A possible compression factor for both XML and Objectivity/DB files can also be included in the comparison.

Another advantage of using the XML file is that the requested data can also be viewed with an XML viewer and the physicist does not have to use a programming language for a quick view of a very small set of data objects. Although a small set of objects can have many associations, these associations can be directly mapped to the XML representation by using logical object identifiers of the object oriented database management system. Thus, it is easy to store all the links and also the linked objects in one XML file.

Since object databases are used at multiple sites possibly spread around the whole world within the CMS collaboration, it is very unlikely that all the analysis centres will have the same hardware. When binary files in the form of Objectivity/DB database files are transferred to such heterogeneous machines, data may not be readable by a particular platform. XML has an outstanding feature for providing data in a format independent of the underlying hardware. Consequently, the XML file which will be produced by extracting data from an object database at platform x can be easily integrated to an object database on platform y.

## 7.3 An Objectivity Query Facility

Currently work in CMS is ongoing in the development of a query facility for the Objectivity database (see [16]). A query facility is a tool, making use of a query language, that interprets a set of user commands constraining the search of data in a database. In relational systems, users make use of the Standard Query Language (SQL), while in object-based systems, the Object Query Language (OQL) is the main query tool. Both SQL and OQL are declarative languages which provide notations for deriving information from relational and object-oriented databases, respectively. In both languages, the user issuing the query must necessarily know the schema of the database to be able to issue a query. In object-oriented databases, the query must know the class names and attributes, at the very least. Consequently, query facilities tend to be domain-driven with the domain schema already loaded into the tool. With different domain schemas, it is very unlikely that the same query facility can be re-used for a different domain.

The Objectivity AS facility is a tool which can help in the creation of a generic query facility for Objectivity databases. A generic query facility for Objectivity databases can use the Objectivity AS facility to load and interrogate the database schema dynamically, thereby not restricting the query facility to a particular domain specification.

XML can facilitate the creation of a generic query facility. The database schema, regardless of the database technology used in creating the schema, can be serialized into a text file, as shown by the work reported in this paper. The query facility can then read the XMLized database schema text file, and can execute queries accordingly. The query facility does not need to make assumptions about the database technology used, as it becomes transparent through the use of XML in providing a standardized language for describing data and objects, and having a common representational view. The generic query facility, with the help of XML and a generic active schema facility, can provide a powerful tool (the Domain Handler, DH [17]) that is not only useful for all domains but is also re-usable. As database querying is an essential functionality in any database management system, the provision and creation of a generic query facility provides a re-usable mechanism for the management and inter-operation of many domains.

The use of a generic query facility, regardless of the domain in use, or the database technology applied, provides a useful guideline in the design of an update facility. An update facility will utilize the generic query facility in the interrogation of the database schema and database elements. The update facility can likewise follow the approach the query facility has taken by not making assumptions on the structure of the database. The update facility should not have any domain-specific semantics within its code. Typically, the update facility invokes the query facility to gain access to the database schema and its objects. For a language- and platform independent database facility, the input and output of the update facility are XML files. This implies that the output of the query facility is serialized into an XML file, i.e. the update facility works with XML strings and not with database objects. Consequently, the output of the update facility is an XML file which passes a Deserializer to transform it back to its object form. Figure 7 illustrates this scheme. Such a setup provides a re-usable mechanism for accessing, interrogating and updating any database system in CMS.

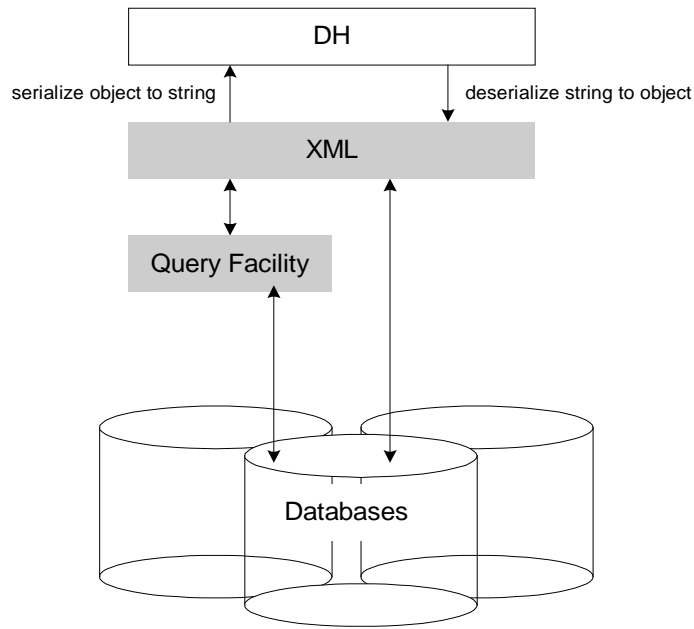

**Figure 7. An XML-Based Generic Query Facility**

## 8. Future Work

The aim of the WISDOM project is to provide general tools which will be used in a number of applications. The initial tools enable data schema from an Objectivity/DB database to have an equivalent representation in XML. Hence it is possible to swap between the two equivalent representations.

This has been extended so that the software can be used to serialize any Objectivity/DB federation and its scope can be reduced to serialize a specific Database, Container or Object within an Objectivity/DB Federation. With some care multiple federations can also be serialized. The following remarks can be made about the design of the software.

## 8.1 Serializable Objects

An object can have the capability to serialize itself, by adding an additional method for this purpose to the object. This is a powerful approach and does not need a schema for producing XML, as the object 'knows' its state and model. Various languages like Java support such objects, which can serialize themselves. Objectivity/DB objects can also be given a serializing capability. However, the current (WISDOM) project is not targeted to any specific database and is rather a general tool for any database. In WISDOM a general approach is being followed that can serialize any type of object from any Objectivity/DB federation.

## 8.2 Selection of the DTD

There can be many possible DTD choices for an Objectivity/DB database. As the objects are inter-linked and contained in containers and databases in an Objectivity/ DB, one could envisage a hierarchical or nested structure in which an object contains all the objects related to it. Alternatively, one can envisage an object in which all container objects are nested and then each container object nests all "basic" objects in it. This may be convenient to give a complete overview of the structure of the database, but if only few objects are taken individually from the XML it will be difficult to identify its position in the database. In WISDOM individual objects are being addressed and each object has its own identity in the XML file and can be individually processed. Also the WISDOM approach does not put any restriction on the scope of the database serialization which may be the case in the alternative approach discussed above.

One cautionary note about DTDs is worthy of inclusion here. In the work reported in this paper, the Objectivity/DB Active Schema API has been used to obtain schema information at run time. So, there is no need to hard code the schema into applications which require schema information in XML. Consequently, the DTD that has been designed is for use with Objectivity/DB schema only. The DTD is not designed for DBMS independence and further DTDs would be required for databases other than Objectivity/DB. This continues to be investigated in the context of the WISDOM project.

## 8.3 Data Compression

One of the main advantages gained by the usage of XML is platform-independent data exchange. However, this comes at the cost of storage. For data intensive transfers over the wide-area network a minimal bandwidth utilization is required and consequently a fast data transfer. [18] has demonstrated that XML files can be compressed very well which yields a reasonable reduction of the files size and thus bandwidth requirements for data transfer. As an example, tests have revealed that an Objectivity database file of 464Kbytes in size can be serialized into an objectXML file of size 21Kbytes and compressed using the Xmill compression tool into a file siz eof only 1.55 Kbytes.

## 9. Conclusions.

In this paper different data interchange needs of the CMS collaboration at CERN have been studied. The fact that such a large collaboration of users is distributed over multiple sites rather than being located at a single site imposes a challenge on data management and integration of data/schema from multiple resources. Based on a commercial object-oriented database management system (Objectivity/DB) a tool has been developed that converts schema and persistent data from Objectivity/DB into a database independent XML format. The tool can be applied to migrate/replicate data sources between remote, distributed sites thus facilitating wide-area database-independent exchange of data.

## Acknowledgements

The authors wish to acknowledge the support of their institutes and, in particular, to thank the numerous developers who have worked on the WISDOM prototypes.